# Acoustic Möbius insulators from projective symmetry

Tianzi Li,[1,†] Juan Du,[1,†] Qicheng Zhang,[1,†] Yitong Li,[1] Xiying Fan,[1]

Fan Zhang,[2,*] and Chunyin Qiu[1,*]

[1]Key Laboratory of Artificial Micro- and Nano-Structures of Ministry of Education and School of Physics and Technology, Wuhan University, Wuhan 430072, China

[2]Department of Physics, University of Texas at Dallas, Richardson, Texas 75080, USA

[†] These authors contributed equally.

[*] To whom correspondence should be addressed: cyqiu@whu.edu.cn; zhang@utdallas.edu

Symmetry plays a critical role in classifying phases of matter. This is exemplified by how crystalline symmetries enrich the topological classification of materials and enable unconventional phenomena in topologically nontrivial ones. After an extensive study over the past decade, the list of topological crystalline insulators and semimetals seems to be exhaustive and concluded. However, in the presence of gauge symmetry, common but not limited to artificial crystals, the algebraic structure of crystalline symmetries needs to be projectively represented, giving rise to unprecedented topological physics. Here we demonstrate this novel idea by exploiting a projective translation symmetry and constructing a variety of Möbius-twisted topological phases. Experimentally, we realize two Möbius insulators in acoustic crystals for the first time: a two-dimensional one of first-order band topology and a three-dimensional one of higher-order band topology. We observe unambiguously the peculiar Möbius edge and hinge states via real-space visualization of their localiztions, momentum-space spectroscopy of their $4\pi$ periodicity, and phase-space winding of their projective translation eigenvalues. Not only does our work open a new avenue for artificial systems under the interplay between gauge and crystalline symmetries, but it also initializes a new framework for topological physics from projective symmetry.

A recurring theme in physics has been the discovery and classification of distinctive phases of matter. In this regard, symmetry and topology are particularly powerful. For instance, the discovery of topological band insulators has taken the research community by storm[1-3]. After the celebrated topological classification[4,5] for the tenfold way of Altland-Zirnbauer symmetry classes including time-reversal, particle-hole, and/or chiral symmetries, the classification has been generalized to systems with spatial symmetries[6-8]. Recently, following the theory of (crystalline) symmetry indicators or topological quantum chemistry, high-throughput screening of topological materials has been performed in the Inorganic Crystal Structure Database, and thousands of candidates have been identified[9-11]. Therefore, the list and classification of topological crystalline phases seem to be exhaustive and concluded. Here we use the translation symmetry to exemplify that, however, in the presence of gauge symmetry[12], the algebraic structure of crystalline symmetries needs to be projectively represented and yields novel topological band physics[13-15]. We first construct theoretically a variety of two- and three-dimensional (2D and 3D), gapped and gapless, topological phases that feature Möbius-twisted boundary states protected by the projective translation symmetry. We then realize experimentally a 2D first-order Möbius insulator (MI) and a 3D higher-order MI (HOMI) for the first time. Particularly, we provide compelling evidence for the projective Möbius topology not only by observing the Möbius edge/hinge states in position, momentum, and energy domains but also in phase domain by revealing the winding of the projective translation eigenvalues. Elucidating the important interplay between gauge and crystalline symmetries, our findings initialize a framework for topological band physics rooted in projective symmetry, given that gauge symmetry is common and abundant in both artificial crystals and interacting systems.

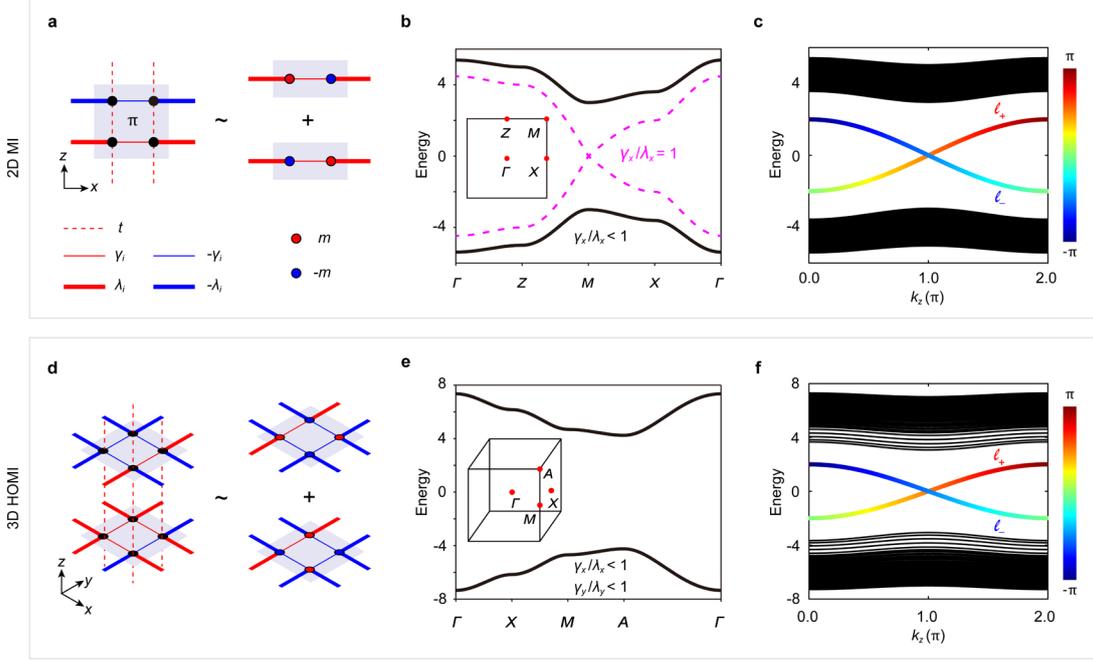

**Figure 1 | 2D first-order Möbius insulator and 3D higher-order Möbius insulator from projective symmetry. a**, Unit cell of the 2D MI (left) and its effective decomposition (right). **b**, Bulk band structures calculated with hoppings $t = 1$ and $\gamma_x = \lambda_x = 2$ (magenta dashed lines), and $t = \gamma_x = 1$ and $\lambda_x = 4$ (black solid lines). Each band is twofold degenerate. **c**, Edge-projected band structure in the $z$ direction for the gapped case in **b** featuring a Möbius twist. **d**, Unit cell of the 3D HOMI (left) and its effective decomposition (right). **e**, Bulk band structure exemplified by a HOMI with hoppings $t = \gamma_x = \gamma_y = 1$ and $\lambda_x = \lambda_y = 4$. Each band is fourfold degenerate. **f**, The hinge-projected band structure in the $z$ direction for the case in **e** featuring a Möbius twist. The sparser black lines are projected surface states. In **c** and **f,** the Möbius twist is formed by two $\pi$-crossed, $4\pi$-periodic, bulk-decoupled bands of opposite projective translation eigenvalues ($\ell_\pm = \pm e^{ik_z/2}$). The color scale indicates the phase profile of $\ell_\pm$.

We start with an elementary 2D model (Fig. 1a) that features projective symmetries[15]. The model Hamiltonian is $H_{2D} = t(1 + \cos k_z)\sigma_0\rho_1 + t\sin k_z\sigma_0\rho_2 + (\gamma_x + \lambda_x\cos k_x)\sigma_1\rho_3 + \lambda_x\sin k_x\sigma_2\rho_3$ $(t, \gamma_x, \lambda_x > 0)$, where $t$ is the hopping in the $z$ direction, $\gamma_x$ and $\lambda_x$ are the intra- and inter-cell hoppings in the $x$ direction, and $\boldsymbol{\sigma}$ and $\boldsymbol{\rho}$ are Pauli matrices acting on the $x$- and $z$-sublattices. As depicted in Fig. 1a, the positive (negative) hoppings are indicated with red (blue) lines, and each plaquette encloses a $\pi$-flux. While $H_{2D}$ does not respect the primitive translation $L_z = \sigma_0 \begin{pmatrix} 0 & 1 \\ e^{ik_z} & 0 \end{pmatrix}$, the inversion $P = \sigma_1 \rho_1$, or the $\mathbb{Z}_2$ gauge transformation $G = \sigma_0 \rho_3$ due to the $\pi$-flux threading, the projective translation $\mathcal{L}_z = GL_z$ and the projective inversion $\mathcal{P} = GP$ are symmetries of the system. Additionally, switching the sublattices $S = \sigma_3 \rho_3$ is a chiral (particle-hole) symmetry of the system. Because of $[\mathcal{L}_z, H_{2D}] = 0$, $H_{2D}$ decouples into two Su-Schrieffer-Heeger (SSH) chains in the $x$ direction of opposite onsite energies $\pm m = \pm 2t\cos(k_z/2)$ and opposite $\mathcal{L}_z$ eigenvalues $\ell_\pm = \pm e^{ik_z/2}$, as illustrated in Fig. 1a. Thus, this system has a $\mathbb{Z}_2$ invariant $\nu$ and is a MI for $\gamma_x < \lambda_x$ (Figs. 1b-1c). Remarkably for the MI phase, the two edge bands in the $k_z$ direction are detached from the bulk and linearly cross at $k_z = \pi$ and $E = 0$ (Fig. 1c), forming a Möbius strip in the edge Brillouin zone and resembling the fractional Josephson effect mediated by two Majorana bound states[16]. The degeneracy at $k_z = \pi$ is Kramers-like and enforced by the projective translation-time symmetry $\mathcal{L}_z T$, since $(\mathcal{L}_z T)^2 = -1$ at $k_z = \pi$. Its pinning to zero energy is a consequence of the chiral symmetry. The two edge states of opposite group velocities are respectively locked with the two $\mathcal{L}_z$ eigenvalues $\ell_\pm$, as indicated in Fig. 1c. As a hallmark of the Möbius topology, $\ell_\pm$ exhibit a $4\pi$ periodicity, again resembling the fractional Josephson effect[16]. Notably, the bulk bands are twofold degenerate (Fig. 1b), reminiscent of the spin-orbit-coupled system with the parity-time symmetry. Our system is spinless, yet the projective algebra $\mathcal{P}^2 = -1$ enforces $(\mathcal{P}T)^2 = -1$ and requires a Kramers degeneracy of the bulk states at every momentum. Significantly, it is the projective inversion symmetry that effectively switches the spinless and spinful nature[17].

Generically, we can extend the 2D model to construct a variety of 3D Möbius phases arising from $\mathbb{Z}_2$ gauge-induced projective symmetries and their algebraic relations. Consider a AB-stacked 3D Hamiltonian, $H_{3D} = \rho_3 h_{2D} + t\sin k_z \rho_2 \mathbb{I} + t(1 + \cos k_z)\rho_1 \mathbb{I}$, where $h_{2D}$ is a 2D monolayer Hamiltonian (replacing the SSH chain above), $\rho$ are Pauli matrices acting on the two layer sublattices, $t$ is the inter-layer hopping, and $\mathbb{I}$ is the identity matrix of the $x$-$y$ plane. Simply, the phase boundary of $H_{3D}$ inherits from that of $h_{2D}$, given that the dispersion relation between the two models: $E_{3D}^2 = E_{2D}^2 + 4t^2\cos^2(k_z/2)$. Next, we show that by appropriately selecting $h_{2D}$, $H_{3D}$ can enjoy gapped and gapless, first-order and higher-order, projective **Möbius** topology.

Figure 1d sketches a model in which the 2D monolayer realizes the quadrupole model[18,19], i.e., $h_{2D} = (\gamma_x + \lambda_x \cos k_x)\tau_0\sigma_1 + \lambda_x \sin k_x \tau_0\sigma_2 + (\gamma_y + \lambda_y \cos k_y)\tau_1\sigma_3 + \lambda_y \sin k_y \tau_2\sigma_3$ ($\gamma_x, \gamma_y, \lambda_x, \lambda_y > 0$), where $\boldsymbol{\sigma}$ and $\boldsymbol{\tau}$ are Pauli matrices acting on the *x*- and *y*-sublattices. The symmetries and their algebraic structures of this 3D model are the same as the 2D MI model. The 3D model has two $\mathbb{Z}_2$ invariants $\tilde{\nu}_x$ and $\tilde{\nu}_y$ (see *Supplementary Information*), and for $(\tilde{\nu}_x, \tilde{\nu}_y) = (1,1)$, i.e., $\gamma_x/\lambda_x<1$ and $\gamma_y/\lambda_y<1$, it realizes a HOMI with protected hinge states. This can be intuitively understood by the fact that in the same parameter regime there exists one protected zero mode per corner per monolayer[18,19]. In this case, the *x-z* and *y-z* surface states are fully gapped, as shown by the sparser black lines in Fig. 1f, yet the hinge states in the $k_z$ direction are Möbius-twisted, as shown by the two crossing lines in Fig. 1f. Because of the projective translation symmetry $[\mathcal{L}_z, H_{3D}] = 0$, $H_{3D}$ decouples into two quadrupole models of opposite $\mathcal{L}_z$ eigenvalues $\ell_\pm = \pm e^{ik_z/2}$ in their topological phases (Fig. 1d). The pair of quadrupole corner states evolves into the π-crossed, 4π-periodic, particle-hole-symmetric hinge states, as enforced by the projective translation-time symmetry $\mathcal{L}_z T$ and the chiral symmetry. Note that our 3D HOMI, which features a Möbius twist in its hinge states, is markedly different from the recently proposed higher-order axion insulator[20], which harbors Möbius surface states and chiral hinge states.

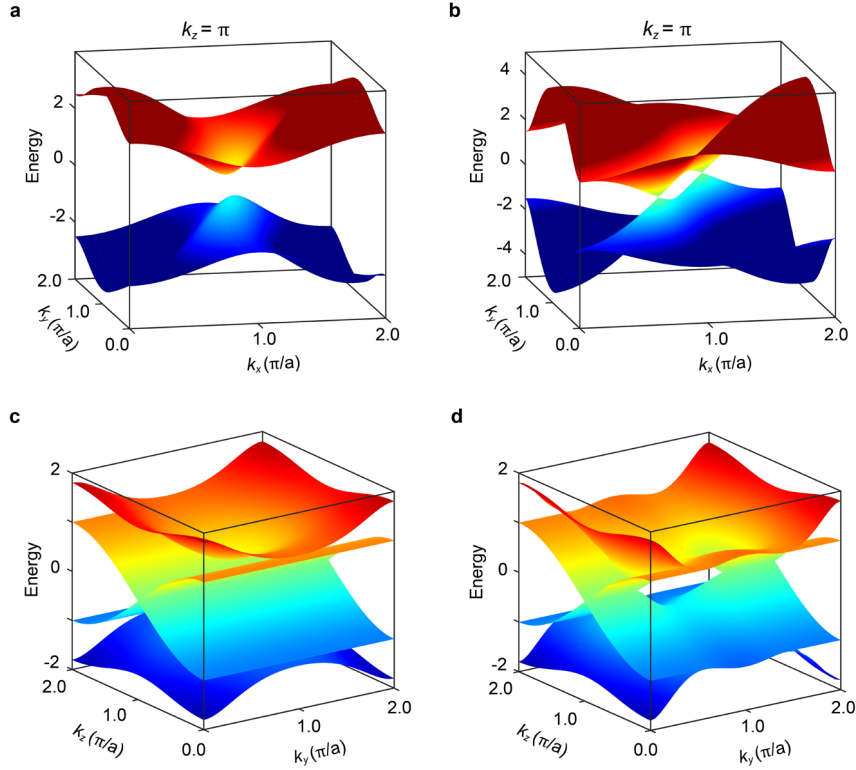

**Figure 2 | 3D first-order Möbius insulator and Möbius Dirac semimetal from projective symmetry. a,b**, Bulk band structures at $k_z = \pi$ exemplified for a 3D MI and a 3D MDS. Each band is twofold degenerate. **c,d**, The corresponding band structures projected into the $k_y$-$k_z$ surface, featuring Möbius-twisted surface states. In **c** the Möbius twist has a zero-energy line degeneracy at $k_z = \pi$ traversing the surface Brillouin zone. In **d** the Möbius twist has a similar line degeneracy but only between the two projected Dirac points.

Figure 2 sketches another 3D model in which the monolayer is a 2D extension of the 1D SSH chain. In its phase diagram there are topological insulator and Dirac semimetal phases (see *Supplementary Information*). Accordingly, the 3D model realizes first-order MI and Möbius Dirac semimetal phases (see *Supplementary Information*). Figs. 2a-2b display clearly their bulk band gap and Dirac points, respectively. In 3D, unlike a nonsymmorphic symmetry that is only invariant at a special 0- or π-momentum plane, here the projective translation symmetry is respected everywhere in the momentum space. Consequently, the surface bands are enforced to exhibit a Möbius *line-twist* instead of a point-twist, as featured in Figs. 2c-2d.

Our Möbius models can be implemented with cavity-tube structures in acoustic systems. Physically, the cavity resonators emulate atomic orbitals, the narrow tubes introduce hoppings between them[21-23], and the tube positions can control the hopping signs[21] to achieve the $\pi$-flux, as visualized in our acoustic crystals (Figs. 3a and 4a). With their structure details in *Supplementary Information*, we have designed a 2D acoustic MI ($t = \gamma_y \approx 68$ Hz, $\lambda_y \approx 261$ Hz, and onsite energy $\approx 5689$ Hz) and a 3D acoustic HOMI ($t \approx 50$ Hz, $\gamma_x = \gamma_y \approx 11$ Hz, $\lambda_x = \lambda_y \approx 157$ Hz, and onsite energy $\approx 5769$ Hz). Both experimental samples are fabricated by 3D printing with a photosensitive resin material, and the fabrication error is of $\sim 0.1$ mm. Next, we experimentally confirm their projective Möbius band topology, not only from the twisted dispersions of edge/hinge states but also from the phase winding of their projective translation eigenvalues.

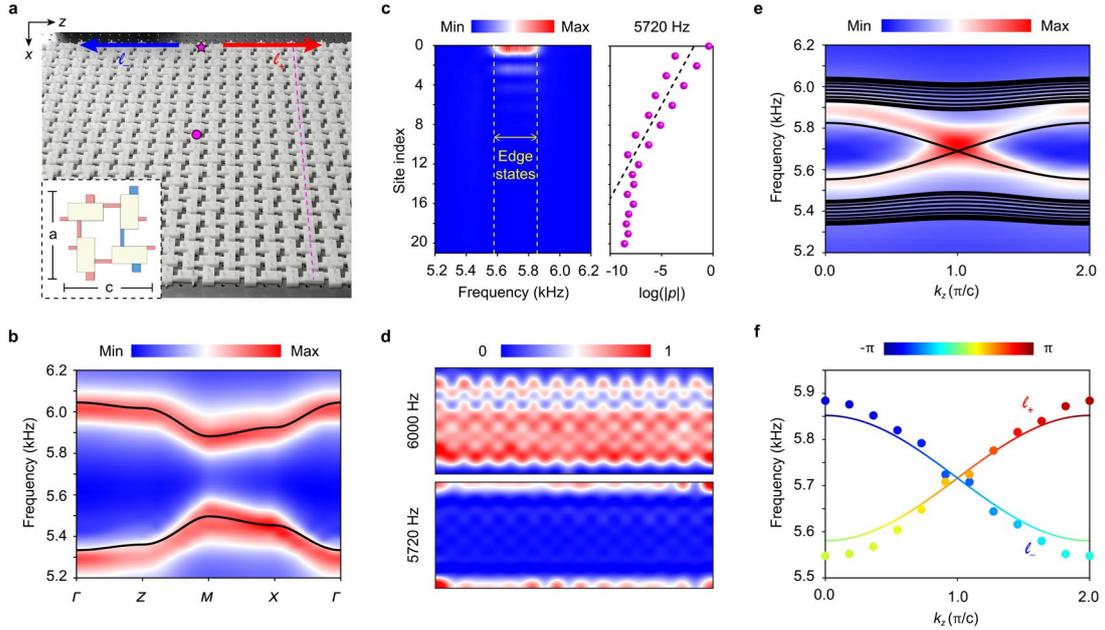

**Figure 3 | Acoustic realization of a 2D Möbius insulator and edge states. a**, Experimental sample. The magenta circle and star label the positions of the sound source in the bulk and edge measurements. Inset: the unit-cell geometry of our acoustic crystal, where the air cavities (white) and narrow tubes (color) mimic the orbitals and hoppings in the tight-binding model, respectively. The lattice constants are $a = c = 75$ mm. **b**, Experimentally measured (color scale) and theoretically predicted (black line) bulk spectra. **c**, Left: frequency-resolved pressure amplitude

scanned along the dashed line in **a**, where the two white dashed lines indicate the frequency window predicted for the edge states. Right: the data extracted at 5720 Hz (magenta spheres) plotted in log scale. **d**, Intensity profiles measured at two selected frequencies respectively for the bulk and edge states. **e**, Measured (color scale) and predicted (black line) edge spectra. **f**, Measured phase information of $\ell_\pm$ (color dots) encoding the measured edge bands, compared with the theoretical results (color lines).

Figure 3a shows our experimental sample for the 2D acoustic MI. It consists of $22 \times 22$ acoustic resonators in the *x* and *y* directions. On each resonator, two small holes were perforated for inserting sound source or probe, and they were sealed when not in use. To measure the bulk band structure, we placed a point-like broadband source in the middle of the sample (magenta circle) and scanned the acoustic response over the sample. Figure 3b presents the Fourier spectrum (color scale) performed for the experimental sound signals in time-space domain. It shows a good agreement with the theoretical prediction (black line). To visualize the edge states, the sound source was relocated to the middle of the top edge (magenta star in Fig. 3a). Figure 3c shows the pressure profile scanned along a row of cavities away from the top edge (dashed line in Fig. 3a). It shows that the sound field is strongly confined to the top edge inside the measured bulk gap and exponentially decays away from the top edge as exemplified at 5720 Hz. The edge states can be further visualized via the spatially resovled acoustic response to local excitations[21,22], as exemplified at 6000 Hz and 5720 Hz for the bulk and edge states (Fig. 3d), respectively, by using a smaller sample of $10 \times 24$ resonators in total.

Figure 3e shows the Fourier spectrum (color scale) performed for the pressure field measured along the top edge. The linear band crossing at $k_z = \pi/c$ provides a direct visualization of the Möbius twist and 4π periodicity in momentum space. The experimental result agrees well with the theoretical prediction (black lines), except for a slight blue shift in frequency (~22 Hz). The band broadening in the experimental data is mainly caused by finite-size effect and unavoidable acoustic dissipation. Intriguingly, because the two edge states of opposite projective translation eigenvalues $\ell_\pm$ have opposite group velocities, they can be distinguished in the left and right regions of the top edge. Moreover, the phase information of $\ell_\pm$ can be extracted from the phase difference of two neighboring sublattices in a *single* unit cell (see *Supplementary Information*). Figure 3f shows our experimentally measured phase evolution of $\ell_\pm$ in the two edge states. Note that the frequencies of color dots correspond to the amplitude peaks in the Fourier spectrum for those given momenta in Fig. 3f. Clearly, it reproduces well the theoretical result (color lines) despite the aforementioned slight blue shift.

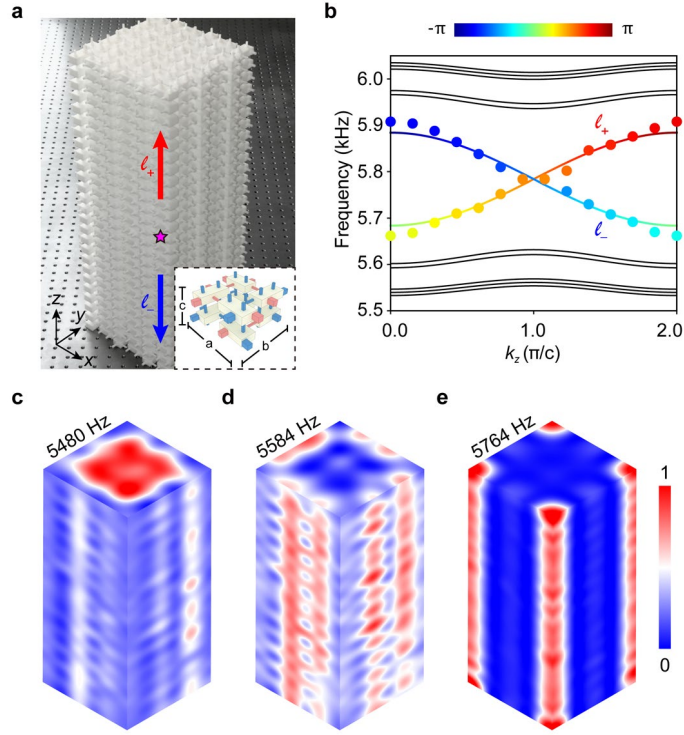

**Figure 4 | Acoustic realziation of a 3D higher-order Möbius insulator and hinge states. a**, Experimental sample, where the inset sketches the unit-cell geometry of our 3D acoustic crystal. The lattice constants are $a = b = 72.6$ mm and $c = 43.6$ mm. **b**, Measured phase information of $\ell_\pm$ (color dots) encoding the measured hinge bands, compared with the theoretical results (color lines). The black lines are projected bulk and surface states. **c-e**, Intensity profiles measured at three selected frequencies respectively for the bulk, surface, and hinge states.

Figure 4a shows our experimental sample for the 3D acoustic HOMI. It consists of $6 \times 6 \times 26$ resonators in the $x$, $y$, and $z$ directions, respectively. A point-like sound source was positioned in the middle of a hinge along the $z$ direction, which excited the hinge states propagating along the $\pm z$ directions simultaneously. Overall, the hinge measurements here were similar to the edge measurements in the 2D case. Figure 4b presents the measured hinge spectrum and the extracted phase information of $\ell_\pm$ for the two hinge bands. Matching well with the theoretical results, they together provide clear evidence for two $\pi$-crossed, $4\pi$-periodic hinge bands. The higher-order band topology can be further visualized via the spatially resovled acoustic response to local excitations[21,22]. In Figs. 4c-4e we show the acoustic intensity fields by sweeping over the sample surfaces for three representative frequencies that are associated to the bulk state (5480 Hz), gapped surface state (5584 Hz), and gapless hinge state (5764 Hz). Note that the gapped surface states appear only at the side surfaces. Compared with Figs. 4c and 4d, Figure 4e shows a strongly hinge-localized sound field, directly demonstrating the presence of higher-order hinge states.

The Möbius twisted boundary states were predicted in fractional Josephson effect[16,24], KHgX (X = As, Sb, Bi)[25], Kondo insulators (CeXSn with X = Ni, Rh, Ir)[26], and axion insulators ($MnBi_{2n}Te_{3n+1}$)[20]. However, they have been elusive in experiment to date[27-29]. For the first time, we provide compelling experimental evidence for the Möbius edge and hinge states in position, momentum, energy, and phase domains. The phase winding is a unique result from the projective translation symmetry. In future, one can fabricate and measure the 3D MI and MDS proposed in Fig. 2, in which the flat Möbius line-twist is another hallmark of the projective translation symmetry. It would also be exciting to explore further the projective inversion symmetry that can switch a spinless system to a spinful one, and vice versa[17]. Having exemplified the interplay between the $\mathbb{Z}_2$ gauge and translation symmetries, our findings urge to establish a complete projective topological classification based on the extraordinarily rich interplay between all variety of gauge and crystalline symmetries, particularly in various artificial systems[30-37] in which gauge symmetries are abundant.

## Methods

**Numerical simulations.**

Acoustic cavity-tube structures were used to mimic our tight-binding model. All full-wave acoustic simulations were performed by using a commercial solver package (COMSOL Multiphysics). The photosensitive resin material used for fabricating samples was modeled as acoustically rigid in the airborne sound environment, given the extremely mismatched acoustic impedance between the resin and air. The air density 1.29 kg/m$^3$ and the sound speed 344.8 m/s were used to solve the eigen-problems (Fig. S5), where the bulk spectra were obtained by imposing the Bloch boundary condition in all directions, and the boundary spectra were simulated by using ribbon structures with the Bloch boundary condition in the edge or hinge direction and the rigid boundary condition in other directions. A detailed fitting process of the hoppings and onsite energies can be referred to *Supplementary Information*.

**Experimental measurements.**

Our experiments were performed for airborne sound at audible frequency. To excite the bulk states, a broadband point-like sound source was located in the middle of the sample, and a 1/4 inch microphone (B&K Type 4187) was used to scan the pressure inside the cavities one by one, together with another identical microphone fixed (at the source cavity) for phase reference. Both the input and output signals were recorded and frequency-resolved with a multi-analyzer system (B&K Type 3560B). The bulk dispersion in Fig. 3b was obtained by performing 2D Fourier transformation of the measured pressure field. The edge spectrum in Fig. 3e was obtained by performing 1D Fourier transform of the measured pressure response along the edge, where the sound source was relocated in the middle of the sample's top edge. The frequencies of the edge bands (Fig. 3f) were extracted from the peaks of the Fourier spectrum at given momenta while their phases $\ell_\pm$ were detailed in *Supplementary Information*, and likewise for the 3D HOMI system in Fig. 4b. In order to make better comparison with the experimental data, tiny frequency shifts of 22 Hz and 15 Hz were introduced to the theoretical results in Fig. 3e and Fig. 4b, respectively. To measure the site-resolved local response (Fig. 3d and Figs. 4c-4e), we placed the source and probe in the same cavity resonator and scanned the acoustic response site by site. Notably, we performed surface measurements to obtain the (projected) information of hinge, surface, and bulk states simultaneously (Figs. 4c-4e).

**Acknowledgements**

We thank M.X. for fruitful discussions. This project is supported by the National Natural Science Foundation of China (Grant No. 11890701, 11674250), the Young Top-Notch Talent for Ten Thousand Talent Program (2019-2022), and the Fundamental Research Funds for the Central Universities (Grant No. 2042020kf0209). F.Z. is supported by the UT Dallas Research Enhancement Fund.

**Author contributions**

C.Q. conceives the idea. T.L. developed the theory and did the simulations. J.D. and Q.Z. designed the experiments and fabricated the samples. J.D., Q.Z., Y.L. and X.F. performed the experiments. C.Q., T.L., J.D., Q.Z. and F.Z. analyzed the data and wrote the manuscript. C.Q. and F.Z. supervised the project. All authors contributed to scientific discussions of the manuscript.

**Author information**

Correspondence and requests for materials should be addressed to C.Q. (cyqiu@whu.edu.cn).

**Reference**

1. Moore, J. E. The birth of topological insulators. *Nature* **464**, 194-198 (2010).
2. Hasan, M. Z. & Kane, C. L. Colloquium: Topological insulators. *Rev. Mod. Phys*. **82**, 3045 (2010).
3. Qi, X.-L. & Zhang, S.-C. Topological insulators and superconductors. *Rev. Mod. Phys*. **83**, 1057 (2011).
4. Kitaev, A. Periodic table for topological insulators and superconductors. *AIP Conf. Proc.* **1134**, 22-30 (2009).
5. Ryu, S., Schnyder, A. P., Furusaki, A. & Ludwig, A. W. W. Topological insulators and superconductors: tenfold way and dimensional hierarchy. *New J. Phys.* **12**, 065010 (2010).
6. Fu, L. Topological crystalline insulators. *Phys. Rev. Lett.* **106**, 106802 (2011).

7. Zhang, F., Kane, C. L., & Mele, E. J. Topological mirror superconductivity. *Phys. Rev. Lett.* **111**, 056403 (2013).

8. Chiu, C. K., Teo, J. C., Schnyder, A. P. & Ryu, S. Classification of topological quantum matter with symmetries. *Rev. Mod. Phys.* **88**, 035005 (2016).

9. Zhang, T., Jiang, Y., Song, Z., Huang, H., He, Y., Fang, Z., Weng, H. & Fang, C. Catalogue of topological electronic materials. *Nature* **566**, 475-479 (2019).

10. Vergniory, M. G., Elcoro, L., Felser, C., Regnault, N., Bernevig, B. A., Wang, Z. A complete catalogue of high-quality topological materials. *Nature* **566**, 480-485 (2019).

11. Tang, F., Po, H. C., Vishwanath, A. & Wan, X. Comprehensive search for topological materials using symmetry indicators. *Nature* **566**, 486-489 (2019).

12. Weinberg, S. The Quantum Theory of Fields (Cambridge University Press, Cambridge, 1995), Vol. 1.

13. Wen, X.-G. Quantum orders and symmetric spin liquids. *Phys. Rev. B* **65**, 165113 (2002).

14. Yang, S. A., Pan, H. & Zhang, F. Dirac and Weyl Superconductors in Three Dimensions. *Phys. Rev. Lett.* **113**, 046401 (2014).

15. Zhao, Y., Huang, Y.-X. & Yang, S. A. $Z_2$-projective translational symmetry protected topological phases. *Phys. Rev. B* **102**, 161117 (2020).

16. Zhang, F. & Pan, W. Fractional Josephson effect: A missing step is a key step. *Nature Materials* **17**, 851-852 (2018).

17. Zhao, Y., Chen, C., Sheng, X.-L. & Yang, S. A. Switching Spinless and Spinful Topological Phases with Projective PT Symmetry, *Phys. Rev. Lett.* **126**, 196402 (2021).

18. Benalcazar, W. A., Bernevig, B. A. & Hughes, T. L. Quantized electric multipole insulators. *Science* **357**, 61-66 (2017).

19. Benalcazar, W. A., Bernevig, B. A. & Hughes, T. L. Electric multipole moments, topological multipole moment pumping, and chiral hinge states in crystalline insulators. *Phys. Rev. B* **96**, 245115 (2017).

20. Zhang, R.-X., Wu, F. & Sarma, S. D. Möbius Insulator and Higher-Order Topology in $MnBi_{2n}Te_{3n+1}$. *Phys. Rev. Lett.* **124**, 136407 (2020).

21. Xue, H. et al. Observation of an acoustic octupole topological insulator. *Nat. Commun.* **11**, 2442 (2020).

22. Ni, X., Li, M., Weiner, M., Alù, A. & Khanikaev, A. B. Demonstration of a quantized acoustic octupole topological insulator. *Nat. Commun.* **11**, 2108 (2020).
23. Qi, Y. et al. Acoustic realization of quadrupole topological insulators. *Phys. Rev. Lett.* **124**, 206601 (2020).
24. Zhang, F. & Kane, C. L. Anomalous topological pumps and fractional Josephson effects. *Phys. Rev. B* **90**, 020501(R) (2014).
25. Wang, Z., Alexandradinata, A., Cava, R. J. & Bernevig, B. A. Hourglass fermions. *Nature* **532**, 189-194 (2016).
26. Chang, P.-Y., Erten, O. & Coleman, P. Möbius kondo insulators. *Nat. Phys.* **13**, 794-798 (2017).
27. Ma, J. et al. Experimental evidence of hourglass fermion in the candidate nonsymmorphic topological insulator KHgSb. *Sci. Adv.* **3**, e1602415 (2017).
28. Liang, A. J. et al. Observation of the topological surface state in the nonsymmorphic topological insulator KHgSb. *Phys. Rev. B* **96**, 165143 (2017).
29. Seong, S. et al. Angle-resolved photoemission spectroscopy study of the Möbius Kondo insulator candidate CeRhSb. *Phys. Rev. B* **100**, 035121 (2019).
30. Lu, L., Joannopoulos, J. D. & Soljačić, M. Topological photonics. *Nat. Photon.* **8**, 821-829 (2014).
31. Ozawa, T. et al. Topological photonics. *Rev. Mod. Phys.* **91**, 015006 (2019).
32. Huber, S. D. Topological mechanics. *Nat. Phys.* **12**, 621-623 (2016).
33. Zhang, X. et al. Topological sound. *Commun. Phys* **1**, 97 (2018).
34. Ma, G., Xiao, M. & Chan, C. T. Topological phases in acoustic and mechanical systems. *Nat. Rev. Phys.* **1**, 281-294 (2019).
35. Xie, B. et al. Higher-order band topology. *Nat. Rev. Phys.* **3**, 520-532 (2021).
36. Goldman, N., Budich, J. C. & Zoller, P. Topological quantum matter with ultracold gases in optical lattices. *Nat. Phys.* **12**, 639-645 (2016).
37. Cooper, N. R., Dalibard, J. & Spielman, I. B. Topological bands for ultracold atoms. *Rev. Mod. Phys.* **91**, 015005 (2019).